\documentclass[
aps,prd,
twocolumn,
preprintnumbers,
amsmath,
amssymb,
nofootinbib,
preprintnumbers
]{revtex4}
\usepackage{here}
\usepackage{footnote}
\usepackage{comment,braket}
\usepackage{epsf}
\usepackage{amsmath}
\usepackage{graphics}
\usepackage{amsfonts}
\usepackage{amssymb}
\usepackage{latexsym}
\usepackage{color}
\usepackage{natbib}
\usepackage{graphicx}
\usepackage{hyperref}
\usepackage{array}

\usepackage[svgnames]{xcolor}
\definecolor{phthaloblue}{rgb}{0.0, 0.06, 0.54}
\hypersetup{
    colorlinks=true,
    linkcolor=blue,
    citecolor=blue,
    filecolor=blue,
    urlcolor=blue,
    }
\begin{document}
\title{Slowly Decaying Ringdown of a Rapidly Spinning Black Hole:\\ Probing the No-Hair Theorem by Small Mass-Ratio Mergers with LISA}

\author{Naritaka Oshita$^{1,2,3}$}
\email{naritaka.oshita@yukawa.kyoto-u.ac.jp}
\author{Daichi Tsuna$^{4,5}$}
\email{tsuna@caltech.edu}
\affiliation{${}^1$RIKEN iTHEMS, Wako, Saitama, 351-0198, Japan}
\affiliation{${}^2$Center for Gravitational Physics and Quantum Information, Yukawa Institute for Theoretical Physics,
Kyoto University, Kitashirakawa Oiwakecho, Sakyo-ku, Kyoto 606-8502, Japan}
\affiliation{${}^3$The Hakubi Center for Advanced Research, Kyoto University,
Yoshida Ushinomiyacho, Sakyo-ku, Kyoto 606-8501, Japan}

\affiliation{
  ${}^4$Research Center for the Early Universe (RESCEU), Graduate School of Science, The University of Tokyo, 7-3-1 Hongo, Bunkyo-ku, Tokyo 113-0033, Japan}

\affiliation{
  ${}^5$TAPIR, Mailcode 350-17, California Institute of Technology, Pasadena, CA 91125, USA}

\preprint{RIKEN-iTHEMS-Report-22}
\preprint{RESCEU-19/22}
\preprint{YITP-23-155}
\begin{abstract}
The measurability of multiple quasinormal (QN) modes, including overtones and higher harmonics, with the Laser Interferometer Space Antenna is investigated by computing the gravitational wave (GW) signal induced by an intermediate or extreme mass ratio merger involving a supermassive black hole (SMBH). We confirm that the ringdown of rapidly spinning black holes are long-lived, and higher harmonics of the ringdown are significantly excited for mergers of small mass ratios. We investigate the measurability and separability of the QN modes for such mergers and demonstrate that the observation of GWs from rapidly rotating SMBHs has an advantage for detecting superposed QN modes and testing the no-hair theorem of black holes.
\end{abstract}

\maketitle

\section{Introduction}
We are in a golden age of gravitational-wave (GW) astronomy, where mergers of binary black holes (BHs) are discovered by GW interferometers \cite{LIGOScientific16,GWTC-1,GWTC-2,GWTC-2.1,GWTC-3,Nitz21,Olsen22}. The end product of a merger is a distorted single BH, which settles down to a Kerr BH by radiating GWs. This {\it ringdown} phase is characterized by a set of damped sinusoids called quasinormal (QN) modes, and is an important probe to test general relativity in the strong-gravity regime \cite{Kokkotas99,Nollert99,Berti09}. QN modes from BH merger remnants have been detected for a large number of events, and were used for various tests of general relativity (e.g., \cite{testGR_150914,testGR_GWTC1,testGR_GWTC2,testGR_GWTC3}).

QN modes consist of fundamental modes and overtones, where the latter is short-lived but can be important for characterizing the ringdown signal \cite{Giesler19,Oshita21}. Detection of overtones from ringdowns is important for e.g. tests of the no-hair theorem \citep{Dreyer:2003bv}. Evidence of an overtone was claimed in the ringdown of GW 150914 \cite{Isi19}, although its significance is still controversial \cite{Cotesta22,Finch22,Isi22}. 

A key parameter that governs the relative strength between fundamental modes and overtones is the spin parameter,\footnote{In this work we use the natural units $c=\hbar=1$ and $G=1$.} $j \equiv J/M^2$, of the remnant BH. $J$ is the angular momentum and $M$ is the mass of the BH. Recently one of the authors found \cite{Oshita21,Oshita:2022pkc} that remnants with rapid spin ($j\gtrsim 0.9$) can have a ringdown dominated by higher overtones and higher angular modes. The more QN modes are detected, the more accurate the test of general relativity would be. Therefore GW ringdown of a highly spinning or near-extremal BH may be a preferred signal to test general relativity. However the final spin of observed BH mergers is typically $\approx 0.7$ \citep{GWTC-1,GWTC-2,GWTC-3}, and such extreme spins may be difficult to probe for mergers of stellar-mass BHs whose natal spins are expected to be rather low \citep{Fuller19}.
\begin{figure*}[t]
  \centering \includegraphics[keepaspectratio=true,height=40mm]{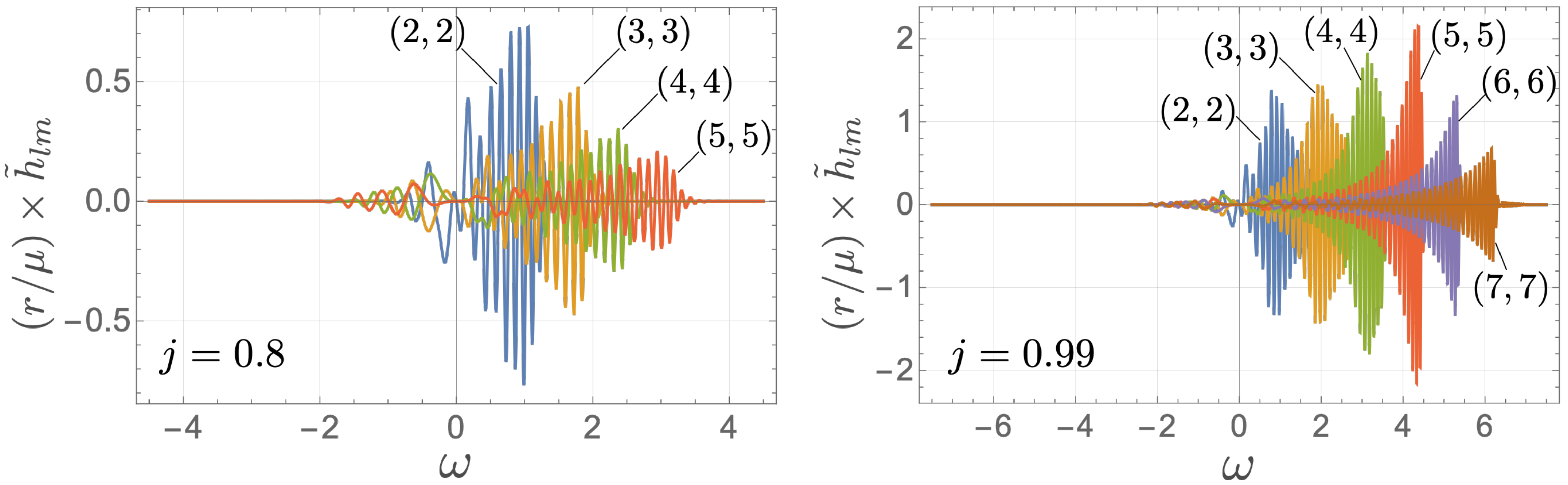}
\caption{Spectra of the GW signal induced by a compact object plunging into a BH with $L_z = 1$ and two BH spins, $j=0.8$ (left) and $0.99$ (right). We adopt the normalization $2M=1$.
}
\centering
\label{spectrum}
\end{figure*}

In this work we consider the possibility of exploring overtones and higher angular modes of rapidly spinning BHs with intermediate/extreme mass ratio mergers involving a supermassive BH (SMBH).\footnote{The self-force of the plunging object is ignored in our computation, as we consider the orbit of a light object. In other words, dephasing of GWs and the backreaction of the object to the trajectory are assumed to be subdominant.} These sources, especially with SMBHs in the mass range $10^6$--$10^7\ M_\odot$, are targets for space-based GW detectors like the Laser Interferometer Space Antenna (LISA) \cite{LISA17,AmaroSeoane22}. Notably these SMBHs are predicted to have large spins due to gas accretion upon their growth (\cite{Dotti13,Dubois14,Bustamante19}, but see \cite{Barausse12}). Although systematic uncertainties may exist in the fitting, X-ray spectra of local SMBHs in this mass range indicate high spins of $>0.9$, consistent with this scenario \citep{Reynolds13,Vasudevan15,Reynolds21}.

Using the waveform modeling of a particle plunging into a rapidly spinning BH and extracting the excited QN modes with a fitting analysis, we estimate the measurability of multiple QN modes, including higher overtones and higher angular modes, by LISA.\footnote{For the LISA detectability of the fundamental mode and the first overtone whose amplitude is assumed to be $1/10$ that of the fundamental one, see Ref. \cite{Berti:2005ys}.} We find that these modes can be detectable out to cosmological distances, realizing a novel probe of gravity in the near-extreme Kerr spacetime. We also evaluate the error of the measurability and separability of the QN modes \cite{Ota:2021ypb,Bhagwat:2021kwv} to assess the feasibility of measuring individual modes.
\section{ringdown for a small mass ratio merger}
\label{sec_ringdown}
In this work we focus on simulating a merger of small mass ratio, such that its dynamics can be well approximated by a test particle plunging into a SMBH. We numerically compute the GW signal induced by a particle plunging into a rotating hole using the Sasaki-Nakamura (SN) equation \cite{Sasaki:1981sx}:
\begin{equation}
\left( \frac{d^2}{dr^{\ast} {}^2} - F_{lm} \frac{d}{dr^{\ast}} -U_{lm} \right) X_{lm} = \tilde{T}_{lm},
\end{equation}
where $X_{lm}$ is a perturbation variable of the gravitational field, $r^{\ast}$ is the tortoise coordinate, $F_{lm}$ and $U_{lm}$ are functions with explicit form given in Ref. \cite{Sasaki:1981sx}, and $\tilde{T}_{lm}$ is the source term associated with the plunging particle.\footnote{To simulate a particle plunging from a finite distance from the SMBH (not from infinity as was assumed in Ref. \cite{Kojima:1984cj}), we modified the source term in Ref. \cite{Kojima:1984cj}. We suppress the contribution of the source term at $\omega \ll 1/M$ including at $\omega =0$ (originating from the particle motion at infinity), by multiplying $f_0$ and $f_1$ in the source term in Appendix B in Ref. \cite{Kojima:1984cj} by $2M \omega$.}
The form of $\tilde{T}_{lm}$ is given in Ref. \cite{Kojima:1984cj} and can be obtained from the geodesic motion of the object. The orbital angular momentum of the plunging orbit is $\mu \times L_z$, and the infalling condition is $-2M (1+\sqrt{1+j}) < L_z < 2 M(1+\sqrt{1-j})$. We here assume that the value of $L_z$ for infalling objects is typically ${\cal O} (M)$ and take $L_z = 2M$ throughout the manuscript.\footnote{We plan to investigate the dependence of GW signals on $L_z$ in a forthcoming paper. Note that when $|L_{\rm max/min} - L_z|/M \ll 1$, where $L_{\rm max}$ and $L_{\rm min}$ are, respectively, the upper and lower limits of $L_z$, the object follows a circulating orbit and the self-force would not be negligible.} Integers $l$ and $m$ are, respectively, the angular and azimuthal numbers of the spheroidal harmonics. We here consider a situation where the trajectory of a compact object of mass $\mu$ is restricted to the equatorial plane\footnote{The Carter constant, one of the parameters characterizing trajectories around BHs, is set to zero in our computation.} ($\theta = \pi/2$) and the total energy of the object (including rest energy) is $\mu$. The self-force of the object can be neglected, which is valid for a small mass ratio $q \equiv \mu/M \ll 1$. Using the Green's function technique, one can solve the SN equation as
\begin{align}
\begin{split}
\lim_{r^{\ast} \to \infty} X_{lm} (\omega,r^{\ast}) &= X_{lm}^{\rm (out)} (\omega) e^{i \omega r^{\ast}}\\
&= \int dr' \tilde{T}_{lm}(r',\omega) G(r',r^{\ast},\omega),
\end{split}
\end{align}
where $G(r',r^{\ast}, \omega)$ is the Green's function that is obtained from the homogeneous solution of the SN equation.
We then obtain the GW spectrum
\begin{equation}
\tilde{h} = \sum_{l,m} \tilde{h}_{lm} (\omega) =\sum_{l,m} - \frac{2}{\omega^2} {}_{-2} S_{lm} (a\omega,\pi/2) R_{lm} (\omega),
\end{equation}
where $a \equiv J/M$ and ${}_{-2} S_{lm}$ is the spin-weighted spheroidal harmonics, assuming an edge-on observer with argument $\pi/2$. The time-domain data $h=h_+ + i h_{\times} = \sum_{(l,m)} h_{lm}$ is obtained by the inverse Laplace transformation of $\tilde{h}$. The function $R_{lm} (\omega)$ is obtained by properly normalizing $X_{lm}$, and its explicit form is provided in Ref. \cite{Sasaki:1981sx}.
\begin{figure}[b]
  \centering \includegraphics[keepaspectratio=true,height=40mm]{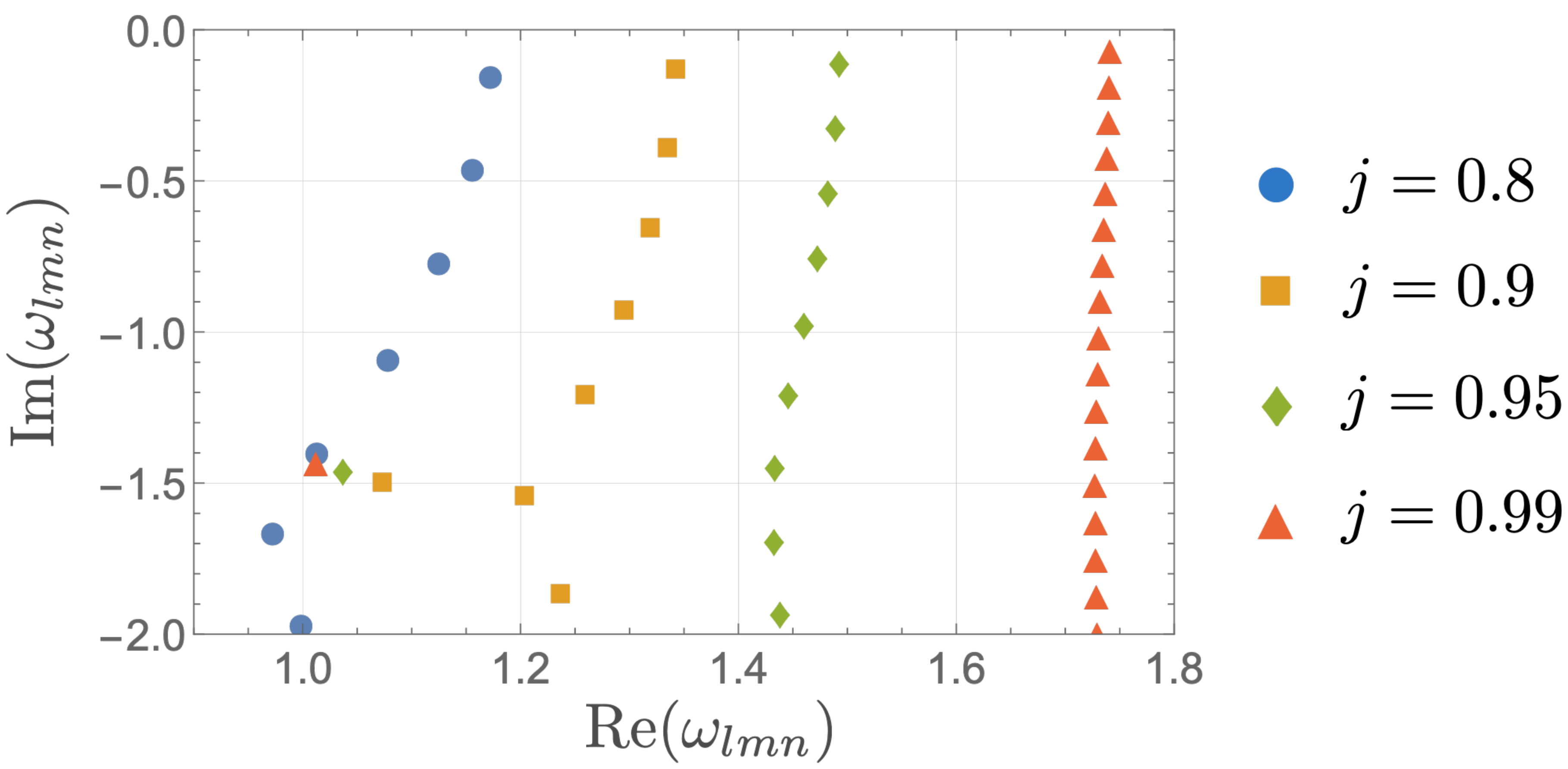}
\caption{Kerr QN frequencies for $(l,m) = (2,2)$, with the normalization $2M=1$. Different markers correspond to different BH spins, and $n$ increases from top to bottom of this plot. For a higher spin, an increasing number of overtones have small Im$(\omega_{lmn})$ and are long-lived.
}
\centering
\label{qnm}
\end{figure}
The GW spectra computed with this scheme is shown in Figure \ref{spectrum}. One can see that $(l,m)=(2,2)$ dominates the GW signal for an intermediate spin ($j=0.8$ in Figure \ref{spectrum}), but higher angular modes are significantly excited for a near-extremal Kerr BH of $j=0.99$.
We are interested in the signal induced by a compact object plunging into a rapidly spinning BH, as more QN modes are long-lived for higher spins (Figure \ref{qnm}).

In the next section, we show that a number of highly damped modes dominate the early ringdown of a near-extremal BH by fitting multiple overtones and fundamental modes to the GW signal. Then we show that rapidly spinning BHs are better targets to perform a high-precision detection of multiple QN modes, including higher angular modes and higher overtones.
\begin{figure}[t]
  \centering \includegraphics[keepaspectratio=true,height=73mm]{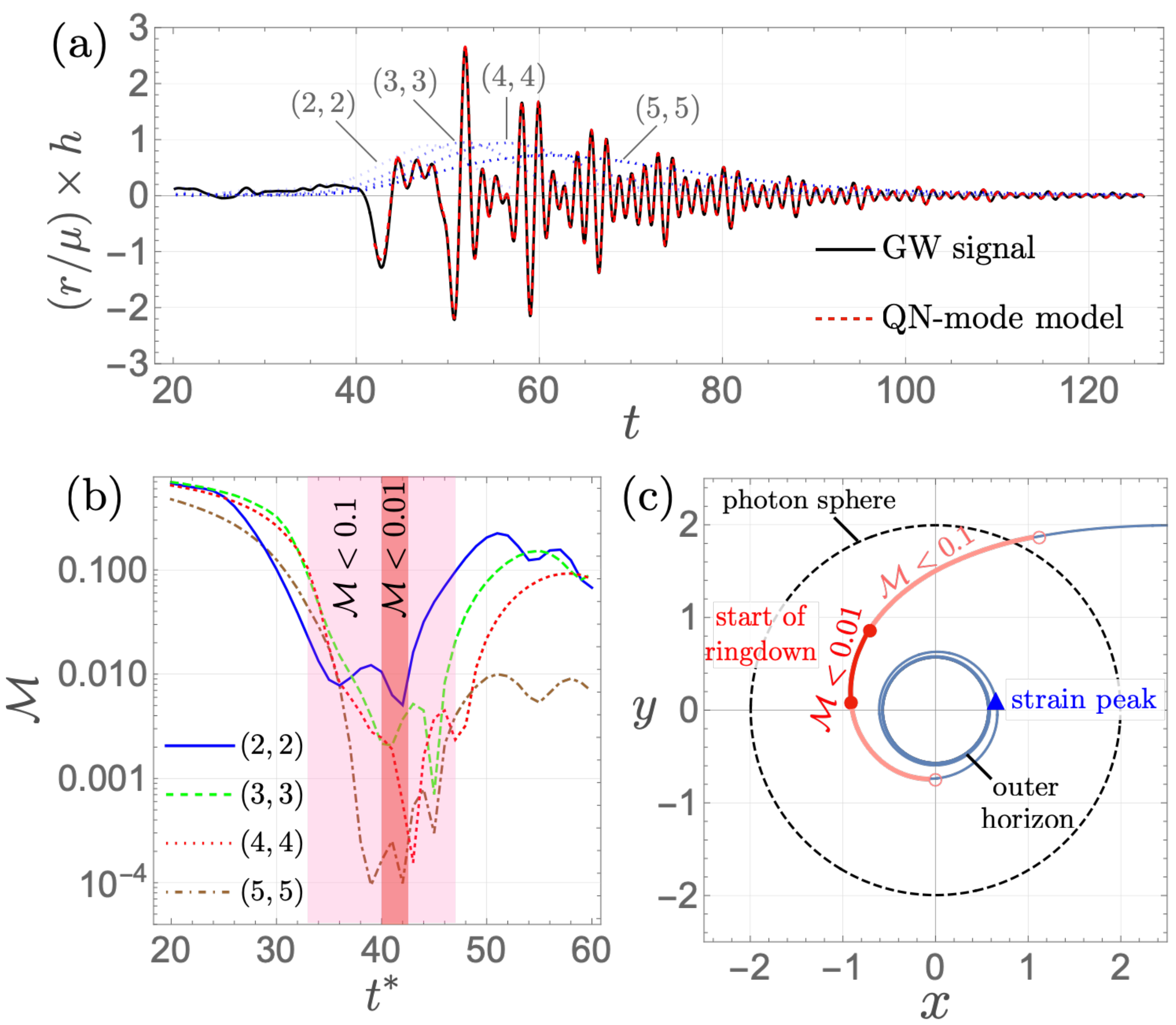}
\caption{Expected GW signal induced by the compact object plunging into a SMBH with $j=0.99$. {\bf (a)} Time-domain waveform of the GW signal $h_{\times}(t)$ (black solid) and the QN-mode model obtained by the fitting analysis, with $t^{\ast} = 42$ and $n_{\rm max} = 20$ for each angular mode in $2 \leq l=m \leq 5$ (red dashed). The blue dotted lines are $|h_{lm}|$ for each angular mode included in our analysis. {\bf (b)} Mismatch ${\cal M}$ with respect to $t^{\ast}$ for $(l,m) = (2,2)$, $(3,3)$, $(4,4)$, and $(5,5)$. All angular modes we computed satisfy ${\cal M}<0.01$ in the range $40 \lesssim t^{\ast} \lesssim 43$.
{\bf (c)} Trajectory of the compact object plunging into the BH. The red and pink lines on the trajectory correspond to the shaded regions in (b) with red (${\cal M} < 0.01$) and pink (${\cal M} < 0.1$), respectively. The blue triangle indicates the position at which the particle sources the signal peak.
}
\centering
\label{trajectory}
\end{figure}
\section{Excitation of overtones}
The measurability of the QN modes is highly sensitive to the start time of ringdown $t^{\ast}$ because (i) it is a superposition of QN modes, each of which is exponentially damped in time, and (ii) overtones may dominate the signal at early times. The exact start time of the ringdown is unknown, but one can obtain a best fit value by fitting multiple QN modes to the GW signal. We then show that the ringdown starts earlier than the strain peak, around the time when the object plunges into the photon sphere. 

\begin{figure}[t]
  \centering \includegraphics[keepaspectratio=true,height=45mm]{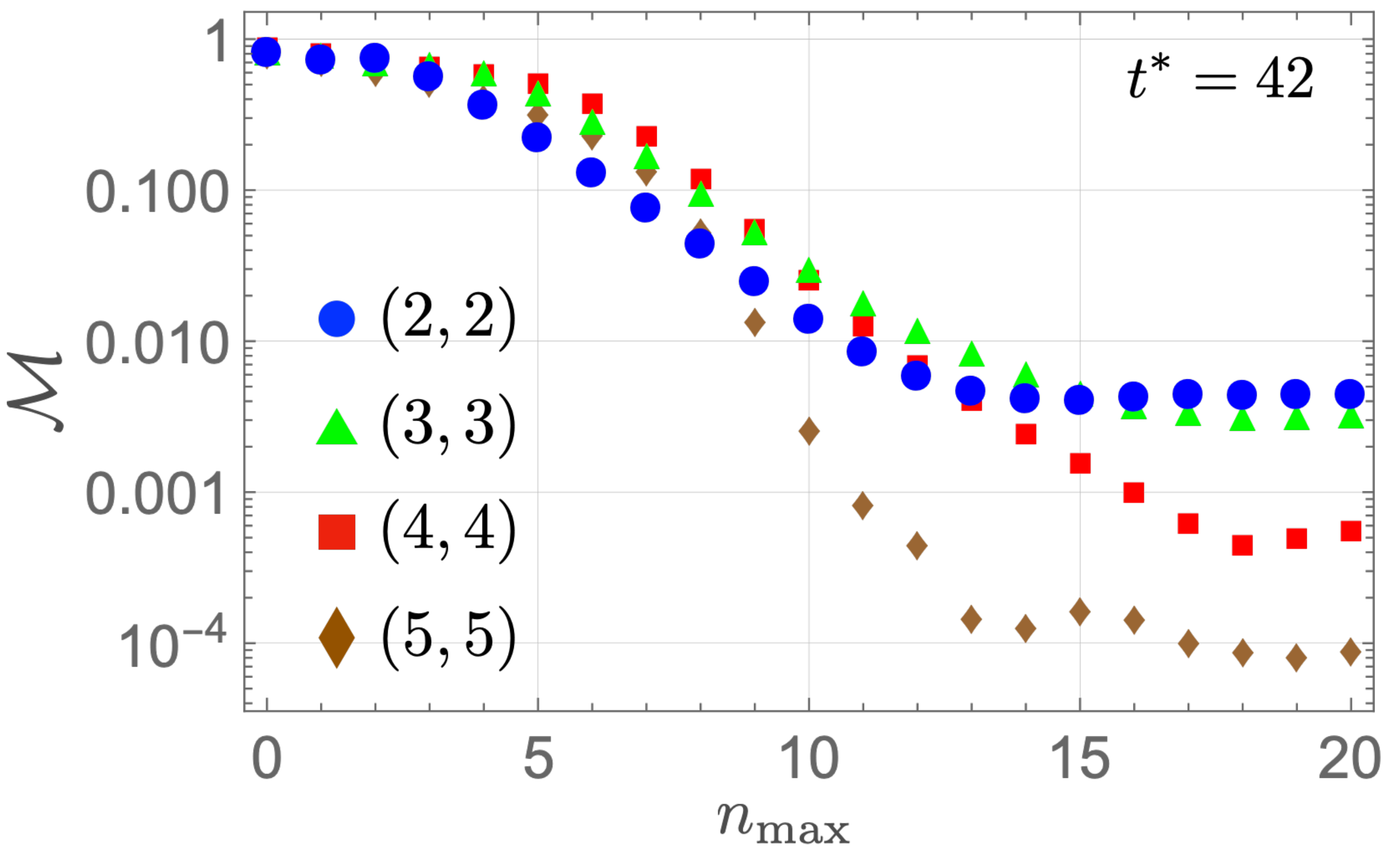}
\caption{Mismatch of the fit of QN modes to the GW data for $j=0.99$. The fitting is performed for $l=m=2$, $3$, $4$, and $5$.
}
\centering
\label{overtones}
\end{figure}

We perform fitting analysis of QN modes\footnote{The pseudospectrum of QN modes implies \cite{Jaramillo:2020tuu} that a small modification in the angular momentum potential or the boundary condition at the horizon may destroy the distribution of the Kerr QN frequencies. However, such an instability could be negligible at the early ringdown (e.g., see Refs. \cite{Cardoso:2016rao,Oshita:2018fqu}), and we are interested in the fit of the standard QN-mode model to the early ringdown in this work.} with a complex frequency $\omega_{lmn}$ including overtones of $n\leq n_{\rm max}$ and angular modes of $2\leq l=m\leq 5$. The fitting is done in the frequency domain by using the QN mode model
\begin{equation}
\tilde{h}_{lm} = \sum_{n}^{n_{\rm max}} \frac{A_{lmn}}{\omega - \omega_{lmn}} e^{i \omega_{lmn} t^{\ast} + i \phi_{lmn}},
\label{spectrum_model}
\end{equation}
to avoid the instability at early ringdown \cite{Oshita:2022pkc} and to treat $t^{\ast}$ as one of the fitting parameters \cite{Finch:2021qph}. In equation (\ref{spectrum_model}), $A_{lmn}$ and $\phi_{lmn}$ are fitting parameters associated with the amplitude and phase of a QN mode with $\omega = \omega_{lmn}$, respectively.
Figure \ref{trajectory}(a) shows that the best fit value of $t^{\ast}$ (in units of $2M$) is in the range $40\lesssim t^{\ast} \lesssim 43$, where the mismatch ${\cal M}$ is less than $0.01$. The range of $t^{\ast}$ with ${\cal M} < 0.1$ corresponds to the moments after the particle plunges into the photon sphere [see Figure \ref{trajectory}(b)]. This is consistent with the fact that GW ringdown is a signal associated with the photon sphere \cite{Ferrari:1984zz}. We also find that including up to higher overtones ($n_{\rm max} \gtrsim 15$) in the fit is necessary to guarantee the convergence of the mismatch (Figure \ref{overtones}). That is, once we admit the start time of ringdown is soon after the compact object plunges into the photon sphere ($t^{\ast} \sim 42$ in our setup), not only the fundamental modes with quadrupole moment but also a number of long-lived modes and higher harmonics are significantly excited for a near-extremal BH. It has an advantage for measuring multiple QN modes, including overtones and higher harmonics, and for accurately testing general relativity. Indeed, it is reported that a fraction of SMBHs can have near-extremal spin parameters of $j \gtrsim 0.99$ \cite{Reynolds13,Vasudevan15,Reynolds21}. In the following, we show such SMBHs are suitable observation targets to detect multiple QN modes.

\section{No-hair test of SMBHs with LISA}
\subsection{Measurability of superposed QN modes}
Let us evaluate the measurability of multiple QN modes of rotating BHs, and its feasibility for tests of the no-hair theorem.
\begin{figure}[t]
  \centering \includegraphics[keepaspectratio=true,height=42mm]{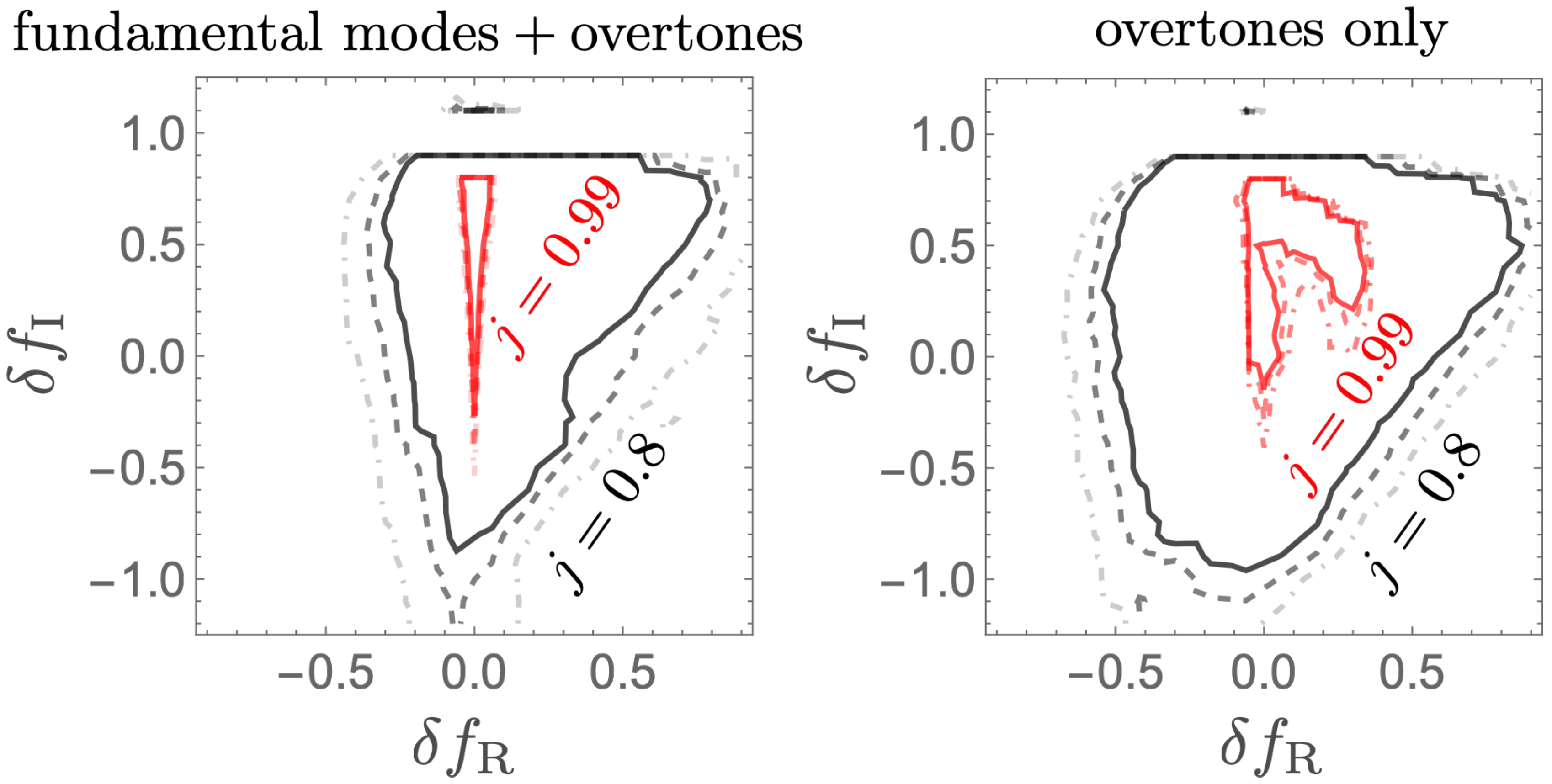}
\caption{Precision for measuring deviations of multiple QN modes from general relativity, for (left) $H_{\delta}^{\rm (F+O)}$ model with $(D_{\rm L}, \tilde{M}, q) = (3\ \text{Gpc}, 10^7 M_{\odot}, 10^{-3})$ and (right) $H_{\delta}^{\rm (O)}$ model with $(1\ \text{Gpc}, 10^7 M_{\odot}, 10^{-3})$. Red and black colors show spins of $j=0.99$ and $0.8$ respectively. Outside the solid, dashed and dot-dashed contours, the likelihood ratio ${\cal L}$ takes values of $> 3.2$, $> 10$, and $>100$, respectively. As higher overtones can be important for higher angular modes (see Figure \ref{overtones} and cf. Ref. \cite{Mourier:2020mwa}), we include many overtones to model the waveforms. For $j=0.99$, we include QN modes up to $n= 21$ for each mode $l=m=2$, $3$, $4$, and $5$. For $j=0.8$, on the other hand, QN modes up to $n= 16$ are included for each mode $l=m=2$, $3$ and $4$.
}
\centering
\label{B_spins}
\end{figure}
We here consider detections of them with LISA, which is sensitive to signals of $\sim 0.01$ Hz and is suitable for detecting ringdown of SMBHs with mass $\sim 10^6$--$10^7 M_{\odot}$. The sensitivity curve can be modeled by an analytic function $\tilde{S}_n (f)$ \cite{Robson:2018ifk}:
\begin{align}
\tilde{S}_n &= \frac{1}{{\cal F} } \left( \frac{P_{\rm OMS}}{L^2} + (1+\cos^2 (f/f_{\ast})) \frac{2 P_{\rm acc}}{(2 \pi f)^4 L^2} \right),\\
{\cal F} &\equiv \frac{3}{10} \left( 1+0.6\times (f/f_{\ast})^2 \right)^{-1},
\end{align}
where $2L$ is the round-trip light travel distance, $f_{\ast} \equiv c/(2 \pi L)$ is the transfer frequency, $P_{\rm OMS}$ and $P_{\rm acc}$ are the single-link optical metrology noise and the single test mass acceleration noise, respectively. The function ${\cal F}$ is the sky/polarization average of the antenna pattern functions.\footnote{Note that the information of the antenna pattern is already included in the noise curve, and we do not need to include this in the signal (see \cite{Robson:2018ifk}).} We include the galactic confusion noise from compact binaries,
\begin{equation}
S_c (f) = A f^{-7/3} e^{-f^{\alpha} +\beta f \sin{(\kappa f)}} [1+\tanh(\gamma (f_k -f))] \text{Hz}^{-1},
\end{equation}
where $A=9 \times 10^{-45}$ and the parameter set $\{ \alpha, \beta, \gamma, \kappa, f_k \}$ is fixed with the values for a four-year mission (we use Table 1 in Ref. \cite{Robson:2018ifk}). Then we obtain the full sensitivity curve as $S_n(f) = \tilde{S}_n (f) +S_c (f)$.
Using $S_n$, we evaluate the SNR of the GW signal and the likelihood ratio, to investigate the support for the model of the no-hair QN modes $H_0$ over a modified model $H_{\delta}$ consisting of a set of complex frequencies deviated from the no-hair values.
\begin{figure*}[t]
  \centering \includegraphics[keepaspectratio=true,height=46mm]{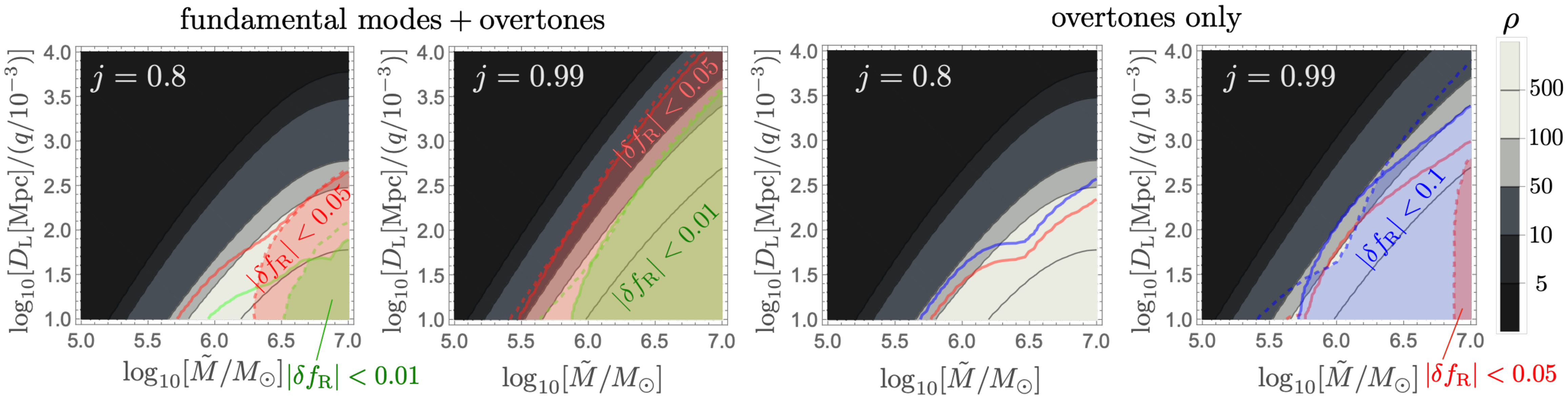}
\caption{Distances out to which tests with multiple QN modes are possible. The contours show the SNR of the ringdown, as a function of the redshifted SMBH mass $\tilde{M}$ and the luminosity distance $D_{\rm L}$ with mass ratio $q$ scaled as $10^{-3}$.
The solid and dashed lines indicate $\delta f_{\rm R} > 0$ and $\delta f_{\rm R} < 0$, respectively. In the shaded regions, below both the solid and dashed lines, one can measure the real part of QN frequencies with (green) $\lesssim 1 \%$, (red) $\lesssim 5 \%$, and (blue) $\lesssim 10\%$, respectively. We here set $\delta f_{\rm I} = 0$, as observations are rather insensitive to $\delta f_{\rm I}$ (see Figure \ref{B_spins}).
}
\centering
\label{BSNR}
\end{figure*}

For the modification of the no-hair model, we consider two types of modifications: (i) a model $H_{\delta}^{\rm (F+O)}$ for which all QN frequencies, including fundamental modes and overtones, are modified as 
\begin{equation}
\omega_{lmn} = \text{Re}(\omega_{lmn}^{\rm (GR)}) (1+\delta f_{\rm R}) + i \text{Im}(\omega_{lmn}^{\rm (GR)}) (1+\delta f_{\rm I}),
\end{equation}
where $\omega_{lmn}^{\rm (GR)}$ are the QN frequencies in general relativity, and (ii) a model $H_{\delta}^{\rm (O)}$ for which only overtones are modified with the above expressions.\footnote{The source parameters of the SMBH ($M$ and $j$) are assumed to be fully known. We leave a more realistic inference with LISA, considering measurement errors of these parameters, to a future study.}
Depending on the parameters of the remnant mass and spin, the whole QN mode frequencies coherently change like the model $H_{\delta}^{\rm (F+O)}$. As such, we can estimate the feasibility of the test of the no-hair theorem, which states that the frequency and decay rate of QN modes are uniquely set by the remnant mass and spin. On the other hand, comparing the likelihood ratio with $H_{\delta}^{\rm (F+O)}$ and the one with $H_{\delta}^{\rm (O)}$, we can see the efficiency of the inclusion of overtones in the test of the no-hair theorem.
The models $H_{0}$, $H_{\delta}$ are respectively given by the superposition of GR or modified QN modes, and the model parameters, i.e., an amplitude and phase, are assigned to each QN mode. The best fit values of the amplitudes and phases are determined by the {\it Mathematica} function ``Fit". Our artificial modification to QN modes affects the model parameters and the likelihood ratio.
The SNR $\rho$ and the likelihood ratio ${\cal L}$ are \cite{Cabero20}
\begin{align}
\rho &= \sqrt{\braket{\tilde{h}_+,\tilde{h}_+} + \braket{\tilde{h}_{\times},\tilde{h}_{\times}}}, \\
{\cal L} &= \frac{p(\tilde{h}_+|H_0)}{p(\tilde{h}_+|H_{\delta})} \frac{p(\tilde{h}_{\times}|H_0)}{p(\tilde{h}_{\times}|H_{\delta})},
\end{align}
where $H_{\delta}$ is $H_{\delta}^{\rm (F+O)}$ or $H_{\delta}^{\rm (O)}$,
\begin{equation}
\braket{x,y} \equiv 4 \text{Re} \int^{\infty}_0 \frac{x(f) y^{\ast} (f)}{S_n} df,
\label{eq:cross_product}
\end{equation}
$\tilde{h}_{+/\times}(f)$ is the Fourier transform of $h_{+/\times} (t)$, and $p(d|H)$ is the likelihood
\begin{equation}
p(d|H(\vec{\vartheta})) \propto \exp \left[ -\frac{1}{2} \braket{d-H(\vec{\vartheta}), d-H(\vec{\vartheta})} \right],
\end{equation}
with a given data, $d$, and a model function, $H(\vec{ \vartheta})$, for a set of the fitting parameters $\vec{\vartheta}$, i.e., the amplitude and phase of each QN mode and $t^{\ast}$.

Figure \ref{B_spins} shows the precision of constraining deviations of multiple QN modes for different thresholds of ${\cal L}$. We find that no-hair tests of rapidly rotating BHs are more powerful than those with BHs of intermediate spin.
For the $H_{\delta}^{\rm (F+O)}$ model in the left panel, the measurement error of the real frequency is $\lesssim 80 \%$ for $j=0.8$, but only $\lesssim 10 \%$ for $j=0.99$. We here take $(D_{\rm L}, \tilde{M},q)=(3~\text{Gpc}, 10^7 M_{\odot}, 10^{-3})$, where $D_{\rm L}$ is the luminosity distance and $\tilde{M} = (1+z) M$ is the redshifted mass of a SMBH that includes the effect of redshift $z$. 

Overtones more quickly damp with time, and hence measuring deviations of overtones is likely more challenging. Nevertheless, for near-extremal BHs, the damping is very weak and even overtones may be measured with good precision. The right panel in Figure \ref{B_spins} shows the high feasibility of measuring overtones for a rapidly spinning BH. We here assume $(D_{\rm L}, \tilde{M}, q) = (1~\text{Gpc}, 10^7 M_{\odot}, 10^{-3})$.
In both models that we considered, the uncertainty in the damping rate $\delta f_{\rm I}$ of QN modes is larger. This would be caused by the dispersive distribution of QN mode frequencies towards the imaginary axis in the complex frequency plane (see Figure \ref{qnm}).
That is, the fundamental mode and overtones have close values of the real part of QN mode frequencies whereas they have dispersive values in the imaginary part. This may cause the large uncertainty in $\delta f_I$ as shown in Figure \ref{B_spins}. A similar conclusion and a large uncertainty in the imaginary part was reported in \cite{Isi19}, where the no-hair test was performed for GW150914 \cite{LIGOScientific:2016aoc}.

Figure \ref{BSNR} summarizes the expected distance out to which we can measure multiple QN modes with high precision. In this section we discuss the prospects for LISA, for moderate and extreme mass ratios.

For moderate mass ratios of $q\sim 10^{-3}$, it corresponds to a merger between a SMBH and an intermediate-mass BH (IMBH). A scenario usually considered for such mergers is clusters hosting IMBHs falling into the galactic nuclei \citep{Miller05,PortegiesZwart06,Matsubayashi07,Arca-Sedda18,Arca-Sedda19}. The event rate is uncertain, but recent N-body simulations find a range $0.003$--$0.03$ Gpc$^{-3}$ yr$^{-1}$ \cite{Arca-Sedda18,Arca-Sedda19}, or $2$--$20$ yr$^{-1}$ within $z<1$ ($D_{\rm L} \lesssim 7$ Gpc) \cite{AmaroSeoane22}. For the $H_{\delta}^{\rm (F+O)}$ model with $j=0.99$ multiple QN frequencies can be measured within $\sim 5\%$ ($\sim 1\%$) for sources at $D_{\rm L}\lesssim 10$ Gpc ($\lesssim 3$ Gpc), and thus no-hair tests of SMBHs are promising. For the $H_{\delta}^{\rm (O)}$ model, one may constrain the real frequencies within $\lesssim 10\%$ for sources out to a few Gpc, corresponding to an event rate of $0.1$--$1$ yr$^{-1}$.

The {likelihood ratio} for the model of $H_{\delta}^{\rm (O)}$ with $\delta f_{\rm R} > 0$ is more significant than that with $\delta f_{\rm R} < 0$ (see Figure \ref{BSNR}). Being sensitive to the modification of $\delta f_{\rm R} > 0$ is reasonable since the higher-frequency modes of $\omega \gtrsim \omega_{lmn}$ in the GW signal are exponentially suppressed (see Ref. \cite{Oshita:2022pkc} for more details).

For extreme mass ratios of $q<10^{-5}$, it corresponds to a stellar-mass BH plunging into a SMBH. Such plunges are expected not to be strong GW emitters, as we also deduce from Figure \ref{BSNR}. The EMRI rate for a Milky-Way like Galaxy is estimated to be $10^{-6}$--$10^{-5}$ yr$^{-1}$, i.e. an event rate of $10^{-8}$--$10^{-7}$ Mpc$^{-3}$ yr$^{-1}$ (\cite{Amaro-Seoane18} and references therein). Plunge orbits can be up to $100$ times more likely than EMRIs \cite{Bar-Or15,Babak18}, so we expect plunges of $\sim 10\ M_\odot$ BHs within the detectable distance ($\lesssim 10$ Mpc) at a rate of $<0.1$ yr$^{-1}$.
Recently a new formation channel of IMBHs in galactic nuclei was proposed, where stellar-mass BHs grow {\it in situ} up to $\sim 10^{4}\ M_\odot$ by collisions with surrounding stars \citep{Rose22}. If such growth is efficient, this would likely enhance the above rates.

\subsection{Separability and measurability of individual QN modes}
In the previous section, we studied the measurability of superposed QN modes, where we required that the SNR for the secondary QN mode is above a given detectability threshold. However, to assess LISA's potential for no-hair tests with ringdown signals, it also is important to evaluate the separability and measurability of individual QN modes (BH spectroscopy).

\begin{figure}[t]
  \centering \includegraphics[keepaspectratio=true,height=85mm]{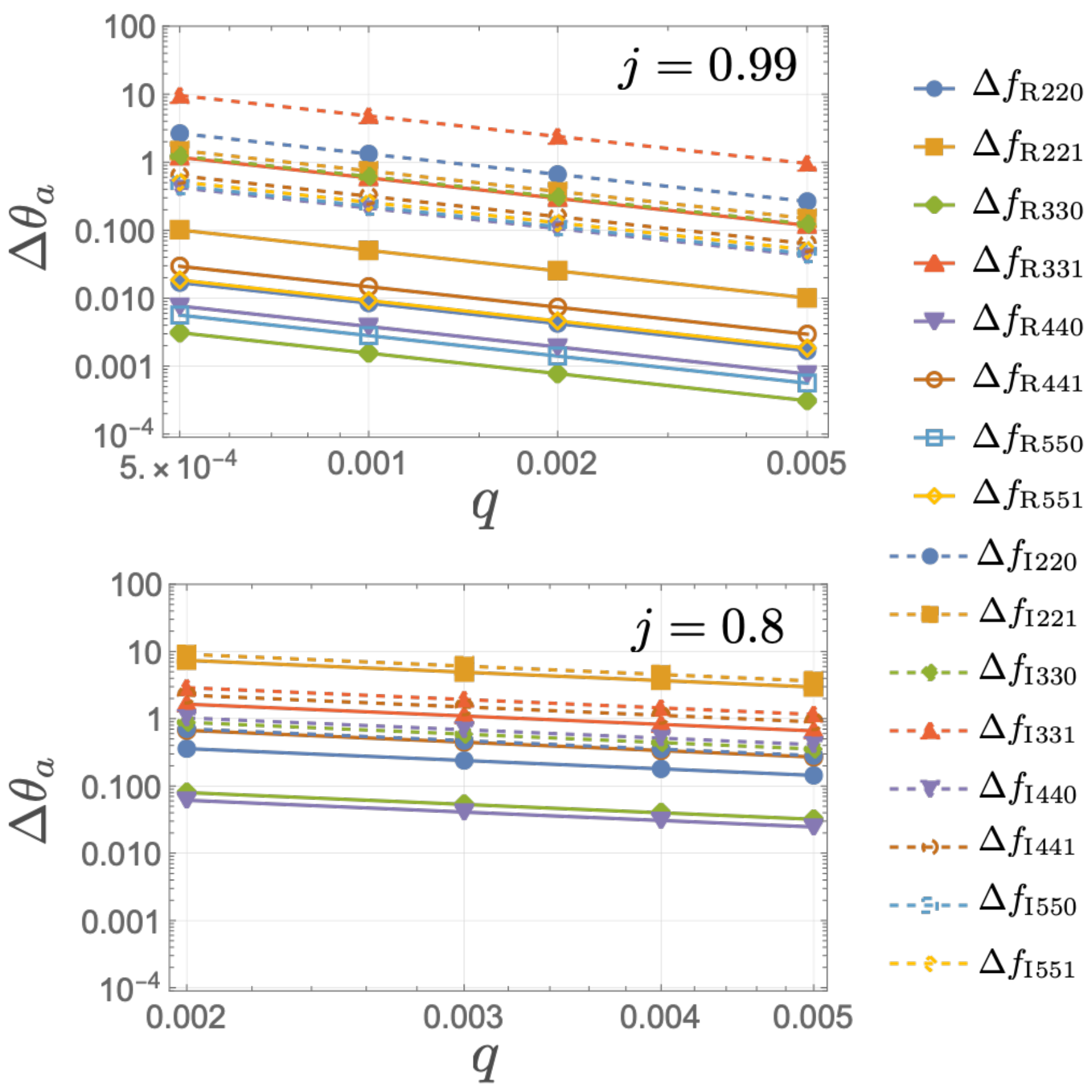}
\caption{The measurability of the real (solid) and imaginary (dashed) parts of the QN modes for a near-extremal $(j=0.99)$ and medium spin $(j=0.8)$ are shown with respect to the mass ratio $q$. The luminosity distance and the mass of the SMBHs are set to $D_{\rm L} = 1$ Gpc and $M = 10^7 M_{\odot}$, respectively.
}
\centering
\label{pig_measurability}
\end{figure}
Let us evaluate the measurability and separability of the fundamental QN mode and the first overtone to see the feasibility of the BH spectroscopy \cite{Ota:2021ypb,Bhagwat:2021kwv}.
The statistical errors on a model parameter $\theta_a$ are given by
\begin{equation}
\sigma_{\theta_{a}} = \sqrt{(\Gamma^{-1})^{aa}}
\end{equation}
where $\Gamma^{-1}$ is the inverse of the Fisher matrix,
\begin{equation}
\Gamma_{ab} = \left\langle \frac{\partial \tilde{h}}{\partial \theta_a} , \frac{\partial \tilde{h}}{\partial \theta_b} \right\rangle,
\label{fismat}
\end{equation}
with $\braket{x,y}$ defined as in equation (\ref{eq:cross_product}).
We here compute the Fisher matrix with the following parameter set
\begin{equation}
\theta_{a} = \bigcup_{lmn} \{f_{\rm R} {}_{lmn}, f_{\rm I} {}_{lmn}, A_{lmn}, \phi_{lmn}\},
\end{equation}
where the parameter set has the fundamental mode ($n=0$) and the first overtone $(n=1)$. The angular modes in the parameter set are $(l,m)=(2,2)$, $(3,3)$, $(4,4)$, $(5,5)$ for $j=0.99$.
For $j=0.8$, it has $(l,m)=(2,2)$, $(3,3)$, $(4,4)$. We use the waveform we numerically obtained in Sec. \ref{sec_ringdown} and use a {\it Mathematica} function ``Fit" to obtain the best fit model of (\ref{spectrum_model}). We then compute the Fisher matrix (\ref{fismat}) by analytically computing the derivative of (\ref{spectrum_model}) and estimate the statistical errors from the inverse matrix $(\Gamma^{-1})_{ab}$. From the statistical errors, we can evaluate the {\it separability} based on the Rayleigh criterion \cite{Ota:2021ypb,Bhagwat:2021kwv}:
\begin{equation}
s[\theta_a,\theta_b] \equiv \text{max} [\sigma_a, \sigma_b] / |\hat{\theta}_a - \hat{\theta}_b|< 1,
\label{separability}
\end{equation}
where $\hat{\theta}_a$ is the true value of $\theta_a$. Also, we can estimate the {\it measurability} (i.e. measurement error) with \cite{Bhagwat:2021kwv}
\begin{equation}
\Delta \theta_a = \sigma_{\theta_a} / \hat{\theta}_a.
\label{measurability}
\end{equation}
The signal has $x\%$ measurability if the set of $\{ \Delta \theta_a\}$ satisfies
\begin{equation}
\max_{i} [\Delta \theta_i] < \frac{x}{100}.
\end{equation}
From this quantity, we can also examine the hierarchy of measurability among the modes we are interested in.

\begin{figure}[t]
  \centering \includegraphics[keepaspectratio=true,height=82mm]{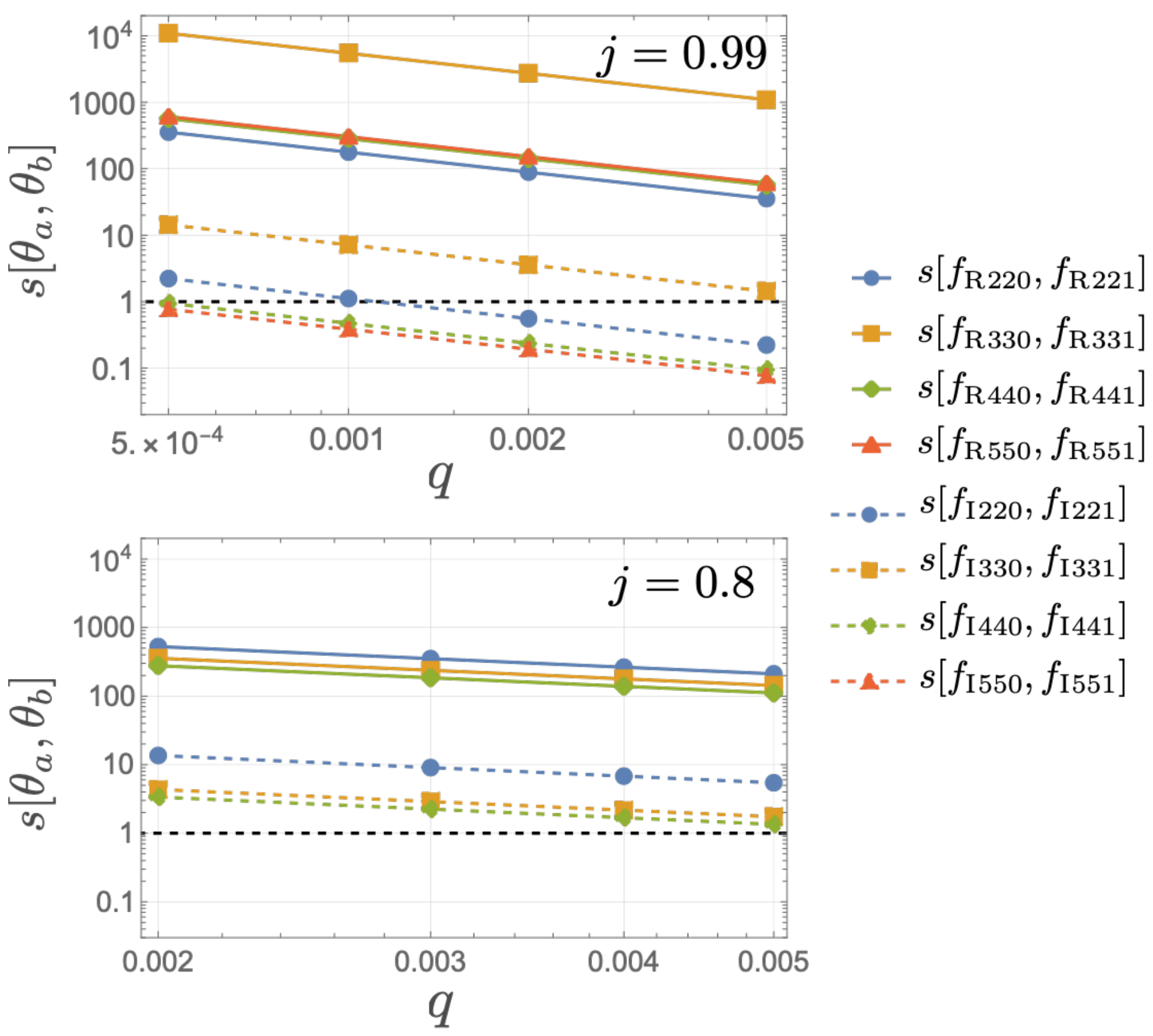}
\caption{
The separability of the real (solid) and imaginary parts of the QN modes for a near-extremal $(j=0.99)$ and medium spin $(j=0.8)$ are shown with respect to the mass ratio $q$. The luminosity distance and the mass of the SMBHs are set to $D_{\rm L} = 1$ Gpc and $M = 10^7 M_{\odot}$, respectively. The black dashed line, below which $s[\theta_a, \theta_b] < 1$, shows the threshold of the separability of $\theta_a$ and $\theta_b$.
}
\centering
\label{pig_separability}
\end{figure}
Figures \ref{pig_measurability} and \ref{pig_separability} show the value of $\Delta {\theta}_a$ and $s [\theta_a, \theta_b]$, respectively. We can read that the errors in the measurability and that in the separability for $j=0.99$ are generally smaller than those for $j=0.8$ at the same luminosity distance $D_{\rm L}$, remnant mass $M$, and the mass ratio of $q \leq 0.005$. The real parts of the QN mode frequencies for $j = 0.8$ all have larger measurement errors. In the case of $j=0.99$, the error of the real part of the QN frequencies with $n=0$ and $(l,m)= (4,4)$ and $(5,5)$ take the smallest values (i.e., highest precision) among them. The real part of QN frequencies of the first overtones can be still measurable in the level of $\Delta f_{{\rm R} {lm1}} \lesssim 0.01$. On the other hand, the imaginary parts of the QN frequencies for higher harmonics are measurable with $\Delta f_{{\rm I} {lm1}} \sim 0.1$.
The error of the imaginary part of QN frequencies in the separability is smaller and can be resolvable for the modes of higher harmonics (see Figure \ref{pig_separability}).
On the other hand, the real parts of QN frequencies for $n=0$ and $n=1$ are too close to resolve especially for $j\sim 0.99$ (see Figures \ref{qnm} and \ref{pig_separability}). The QN modes for $j=0.8$ are difficult to distinguish each other in our setup as the damping rates in QN modes are larger.

In a previous work \cite{Bhagwat:2021kwv}, such measurement errors by LISA were computed for mergers of nonspinning BHs with mass ratio of $0.1 \lesssim q \leq 1$ and total mass of $10^6M_\odot$. While varying $q$ just changes the overall scale of the measurement error, varying the remnant mass and spin may change even the hierarchy among different modes. Indeed, our result shows that the higher angular modes, i.e., $(4,4,0)$ and $(5,5,0)$, have the first three smallest measurement errors for $j=0.99$ whereas $(2,2,0)$ mode takes the smallest error for the case considered by \cite{Bhagwat:2021kwv} (their Figure 4). As the higher angular modes may dominate the ringdown signal for a rapidly spinning BH as shown in Figure \ref{spectrum}, the rapid spin of the remnant BH may affect the hierarchy.

\section{Conclusion}
In this paper, we studied the measurability and separability of multiple QN modes emitted by near extremal SMBHs, which may exist at the center of galaxies according to the X-ray observation of the accretion disks \cite{Dotti13,Dubois14,Bustamante19}.
The measurability of superposed QN modes is estimated by the SNR of a ringdown signal that is obtained by the fit of QN modes to the whole GW data (Figure \ref{BSNR}). The goodness of the fit with the GR QN modes was assessed by the likelihood ratio (Figures \ref{B_spins} and \ref{BSNR}). To assess the ability of the BH spectroscopy, we computed the statistical error to obtain the errors in the separability (\ref{separability}) and in the measurability (\ref{measurability}). We then found that the separability and measurability for mergers involving near-extremal SMBHs of $j=0.99$ are generally better than those with SMBHs of moderate spins of $j=0.8$ (Figures \ref{pig_measurability} and \ref{pig_separability}). The measurement error of the real part of QN frequencies can be $\lesssim 1\%$ and the separability condition is satisfied for the imaginary part when $D_{\rm L} \lesssim 1$ Gpc and $q\sim 0.005$.
We thus conclude that intermediate (and possibly extreme) mass ratio mergers can be unique targets for LISA to probe multiple QN modes of rapidly spinning BHs, and an important target for tests of gravity in a near-extreme Kerr spacetime. 

\begin{acknowledgements}
N. O. was supported by the Special Postdoctoral Researcher (SPDR) Program at RIKEN, FY2021 Incentive Research Project at RIKEN, Grant-in-Aid for Scientific Research (KAKENHI) project for FY 2021 (JP21K20371) and FY2023 (JP23K13111). D.T. is supported by the Sherman Fairchild Postdoctoral Fellowship at
the California Institute of Technology.
\end{acknowledgements}


\begin{thebibliography}{55}
  \expandafter\ifx\csname natexlab\endcsname\relax\def\natexlab#1{#1}\fi
  \expandafter\ifx\csname bibnamefont\endcsname\relax
    \def\bibnamefont#1{#1}\fi
  \expandafter\ifx\csname bibfnamefont\endcsname\relax
    \def\bibfnamefont#1{#1}\fi
  \expandafter\ifx\csname citenamefont\endcsname\relax
    \def\citenamefont#1{#1}\fi
  \expandafter\ifx\csname url\endcsname\relax
    \def\url#1{\texttt{#1}}\fi
  \expandafter\ifx\csname urlprefix\endcsname\relax\def\urlprefix{URL }\fi
  \providecommand{\bibinfo}[2]{#2}
  \providecommand{\eprint}[2][]{\url{#2}}
  
  \bibitem[{\citenamefont{Abbott et~al.}(2016{\natexlab{a}})}]{LIGOScientific16}
  \bibinfo{author}{\bibfnamefont{B.~P.} \bibnamefont{Abbott}} \bibnamefont{et~al.} (\bibinfo{collaboration}{LIGO Scientific, Virgo}), \bibinfo{journal}{Phys. Rev. X} \textbf{\bibinfo{volume}{6}}, \bibinfo{pages}{041015} (\bibinfo{year}{2016}{\natexlab{a}}), \bibinfo{note}{[Erratum: Phys.Rev.X 8, 039903 (2018)]}, \eprint{1606.04856}.
  
  \bibitem[{\citenamefont{Abbott et~al.}(2019{\natexlab{a}})}]{GWTC-1}
  \bibinfo{author}{\bibfnamefont{B.~P.} \bibnamefont{Abbott}} \bibnamefont{et~al.} (\bibinfo{collaboration}{LIGO Scientific, Virgo}), \bibinfo{journal}{Phys. Rev. X} \textbf{\bibinfo{volume}{9}}, \bibinfo{pages}{031040} (\bibinfo{year}{2019}{\natexlab{a}}), \eprint{1811.12907}.
  
  \bibitem[{\citenamefont{Abbott et~al.}(2021{\natexlab{a}})}]{GWTC-2}
  \bibinfo{author}{\bibfnamefont{R.}~\bibnamefont{Abbott}} \bibnamefont{et~al.} (\bibinfo{collaboration}{LIGO Scientific, Virgo}), \bibinfo{journal}{Phys. Rev. X} \textbf{\bibinfo{volume}{11}}, \bibinfo{pages}{021053} (\bibinfo{year}{2021}{\natexlab{a}}), \eprint{2010.14527}.
  
  \bibitem[{\citenamefont{Abbott et~al.}(2021{\natexlab{b}})}]{GWTC-2.1}
  \bibinfo{author}{\bibfnamefont{R.}~\bibnamefont{Abbott}} \bibnamefont{et~al.} (\bibinfo{collaboration}{LIGO Scientific, VIRGO}) (\bibinfo{year}{2021}{\natexlab{b}}), \eprint{2108.01045}.
  
  \bibitem[{\citenamefont{Abbott et~al.}(2021{\natexlab{c}})}]{GWTC-3}
  \bibinfo{author}{\bibfnamefont{R.}~\bibnamefont{Abbott}} \bibnamefont{et~al.} (\bibinfo{collaboration}{LIGO Scientific, VIRGO, KAGRA}) (\bibinfo{year}{2021}{\natexlab{c}}), \eprint{2111.03606}.
  
  \bibitem[{\citenamefont{Nitz et~al.}(2021)\citenamefont{Nitz, Capano, Kumar, Wang, Kastha, Sch\"afer, Dhurkunde, and Cabero}}]{Nitz21}
  \bibinfo{author}{\bibfnamefont{A.~H.} \bibnamefont{Nitz}}, \bibinfo{author}{\bibfnamefont{C.~D.} \bibnamefont{Capano}}, \bibinfo{author}{\bibfnamefont{S.}~\bibnamefont{Kumar}}, \bibinfo{author}{\bibfnamefont{Y.-F.} \bibnamefont{Wang}}, \bibinfo{author}{\bibfnamefont{S.}~\bibnamefont{Kastha}}, \bibinfo{author}{\bibfnamefont{M.}~\bibnamefont{Sch\"afer}}, \bibinfo{author}{\bibfnamefont{R.}~\bibnamefont{Dhurkunde}}, \bibnamefont{and} \bibinfo{author}{\bibfnamefont{M.}~\bibnamefont{Cabero}}, \bibinfo{journal}{Astrophys. J.} \textbf{\bibinfo{volume}{922}}, \bibinfo{pages}{76} (\bibinfo{year}{2021}), \eprint{2105.09151}.
  
  \bibitem[{\citenamefont{Olsen et~al.}(2022)\citenamefont{Olsen, Venumadhav, Mushkin, Roulet, Zackay, and Zaldarriaga}}]{Olsen22}
  \bibinfo{author}{\bibfnamefont{S.}~\bibnamefont{Olsen}}, \bibinfo{author}{\bibfnamefont{T.}~\bibnamefont{Venumadhav}}, \bibinfo{author}{\bibfnamefont{J.}~\bibnamefont{Mushkin}}, \bibinfo{author}{\bibfnamefont{J.}~\bibnamefont{Roulet}}, \bibinfo{author}{\bibfnamefont{B.}~\bibnamefont{Zackay}}, \bibnamefont{and} \bibinfo{author}{\bibfnamefont{M.}~\bibnamefont{Zaldarriaga}} (\bibinfo{collaboration}{LIGO Scientific Collaboration, the Virgo}), \bibinfo{journal}{Phys. Rev. D} \textbf{\bibinfo{volume}{106}}, \bibinfo{pages}{043009} (\bibinfo{year}{2022}), \eprint{2201.02252}.
  
  \bibitem[{\citenamefont{Kokkotas and Schmidt}(1999)}]{Kokkotas99}
  \bibinfo{author}{\bibfnamefont{K.~D.} \bibnamefont{Kokkotas}} \bibnamefont{and} \bibinfo{author}{\bibfnamefont{B.~G.} \bibnamefont{Schmidt}}, \bibinfo{journal}{Living Rev. Rel.} \textbf{\bibinfo{volume}{2}}, \bibinfo{pages}{2} (\bibinfo{year}{1999}), \eprint{gr-qc/9909058}.
  
  \bibitem[{\citenamefont{Nollert}(1999)}]{Nollert99}
  \bibinfo{author}{\bibfnamefont{H.-P.} \bibnamefont{Nollert}}, \bibinfo{journal}{Class. Quant. Grav.} \textbf{\bibinfo{volume}{16}}, \bibinfo{pages}{R159} (\bibinfo{year}{1999}).
  
  \bibitem[{\citenamefont{Berti et~al.}(2009)\citenamefont{Berti, Cardoso, and Starinets}}]{Berti09}
  \bibinfo{author}{\bibfnamefont{E.}~\bibnamefont{Berti}}, \bibinfo{author}{\bibfnamefont{V.}~\bibnamefont{Cardoso}}, \bibnamefont{and} \bibinfo{author}{\bibfnamefont{A.~O.} \bibnamefont{Starinets}}, \bibinfo{journal}{Class. Quant. Grav.} \textbf{\bibinfo{volume}{26}}, \bibinfo{pages}{163001} (\bibinfo{year}{2009}), \eprint{0905.2975}.
  
  \bibitem[{\citenamefont{Abbott et~al.}(2016{\natexlab{b}})}]{testGR_150914}
  \bibinfo{author}{\bibfnamefont{B.~P.} \bibnamefont{Abbott}} \bibnamefont{et~al.} (\bibinfo{collaboration}{LIGO Scientific, Virgo}), \bibinfo{journal}{Phys. Rev. Lett.} \textbf{\bibinfo{volume}{116}}, \bibinfo{pages}{221101} (\bibinfo{year}{2016}{\natexlab{b}}), \bibinfo{note}{[Erratum: Phys.Rev.Lett. 121, 129902 (2018)]}, \eprint{1602.03841}.
  
  \bibitem[{\citenamefont{Abbott et~al.}(2019{\natexlab{b}})}]{testGR_GWTC1}
  \bibinfo{author}{\bibfnamefont{B.~P.} \bibnamefont{Abbott}} \bibnamefont{et~al.} (\bibinfo{collaboration}{LIGO Scientific, Virgo}), \bibinfo{journal}{Phys. Rev. D} \textbf{\bibinfo{volume}{100}}, \bibinfo{pages}{104036} (\bibinfo{year}{2019}{\natexlab{b}}), \eprint{1903.04467}.
  
  \bibitem[{\citenamefont{Abbott et~al.}(2021{\natexlab{d}})}]{testGR_GWTC2}
  \bibinfo{author}{\bibfnamefont{R.}~\bibnamefont{Abbott}} \bibnamefont{et~al.} (\bibinfo{collaboration}{LIGO Scientific, Virgo}), \bibinfo{journal}{Phys. Rev. D} \textbf{\bibinfo{volume}{103}}, \bibinfo{pages}{122002} (\bibinfo{year}{2021}{\natexlab{d}}), \eprint{2010.14529}.
  
  \bibitem[{\citenamefont{Abbott et~al.}(2021{\natexlab{e}})}]{testGR_GWTC3}
  \bibinfo{author}{\bibfnamefont{R.}~\bibnamefont{Abbott}} \bibnamefont{et~al.} (\bibinfo{collaboration}{LIGO Scientific, VIRGO, KAGRA}) (\bibinfo{year}{2021}{\natexlab{e}}), \eprint{2112.06861}.
  
  \bibitem[{\citenamefont{Giesler et~al.}(2019)\citenamefont{Giesler, Isi, Scheel, and Teukolsky}}]{Giesler19}
  \bibinfo{author}{\bibfnamefont{M.}~\bibnamefont{Giesler}}, \bibinfo{author}{\bibfnamefont{M.}~\bibnamefont{Isi}}, \bibinfo{author}{\bibfnamefont{M.~A.} \bibnamefont{Scheel}}, \bibnamefont{and} \bibinfo{author}{\bibfnamefont{S.}~\bibnamefont{Teukolsky}}, \bibinfo{journal}{Phys. Rev. X} \textbf{\bibinfo{volume}{9}}, \bibinfo{pages}{041060} (\bibinfo{year}{2019}), \eprint{1903.08284}.
  
  \bibitem[{\citenamefont{Oshita}(2021)}]{Oshita21}
  \bibinfo{author}{\bibfnamefont{N.}~\bibnamefont{Oshita}}, \bibinfo{journal}{Phys. Rev. D} \textbf{\bibinfo{volume}{104}}, \bibinfo{pages}{124032} (\bibinfo{year}{2021}), \eprint{2109.09757}.
  
  \bibitem[{\citenamefont{Dreyer et~al.}(2004)\citenamefont{Dreyer, Kelly, Krishnan, Finn, Garrison, and Lopez-Aleman}}]{Dreyer:2003bv}
  \bibinfo{author}{\bibfnamefont{O.}~\bibnamefont{Dreyer}}, \bibinfo{author}{\bibfnamefont{B.~J.} \bibnamefont{Kelly}}, \bibinfo{author}{\bibfnamefont{B.}~\bibnamefont{Krishnan}}, \bibinfo{author}{\bibfnamefont{L.~S.} \bibnamefont{Finn}}, \bibinfo{author}{\bibfnamefont{D.}~\bibnamefont{Garrison}}, \bibnamefont{and} \bibinfo{author}{\bibfnamefont{R.}~\bibnamefont{Lopez-Aleman}}, \bibinfo{journal}{Class. Quant. Grav.} \textbf{\bibinfo{volume}{21}}, \bibinfo{pages}{787} (\bibinfo{year}{2004}), \eprint{gr-qc/0309007}.
  
  \bibitem[{\citenamefont{Isi et~al.}(2019)\citenamefont{Isi, Giesler, Farr, Scheel, and Teukolsky}}]{Isi19}
  \bibinfo{author}{\bibfnamefont{M.}~\bibnamefont{Isi}}, \bibinfo{author}{\bibfnamefont{M.}~\bibnamefont{Giesler}}, \bibinfo{author}{\bibfnamefont{W.~M.} \bibnamefont{Farr}}, \bibinfo{author}{\bibfnamefont{M.~A.} \bibnamefont{Scheel}}, \bibnamefont{and} \bibinfo{author}{\bibfnamefont{S.~A.} \bibnamefont{Teukolsky}}, \bibinfo{journal}{Phys. Rev. Lett.} \textbf{\bibinfo{volume}{123}}, \bibinfo{pages}{111102} (\bibinfo{year}{2019}), \eprint{1905.00869}.
  
  \bibitem[{\citenamefont{Cotesta et~al.}(2022)\citenamefont{Cotesta, Carullo, Berti, and Cardoso}}]{Cotesta22}
  \bibinfo{author}{\bibfnamefont{R.}~\bibnamefont{Cotesta}}, \bibinfo{author}{\bibfnamefont{G.}~\bibnamefont{Carullo}}, \bibinfo{author}{\bibfnamefont{E.}~\bibnamefont{Berti}}, \bibnamefont{and} \bibinfo{author}{\bibfnamefont{V.}~\bibnamefont{Cardoso}} (\bibinfo{year}{2022}), \eprint{2201.00822}.
  
  \bibitem[{\citenamefont{Finch and Moore}(2022)}]{Finch22}
  \bibinfo{author}{\bibfnamefont{E.}~\bibnamefont{Finch}} \bibnamefont{and} \bibinfo{author}{\bibfnamefont{C.~J.} \bibnamefont{Moore}}, \bibinfo{journal}{Phys. Rev. D} \textbf{\bibinfo{volume}{106}}, \bibinfo{pages}{043005} (\bibinfo{year}{2022}), \eprint{2205.07809}.
  
  \bibitem[{\citenamefont{Isi and Farr}(2022)}]{Isi22}
  \bibinfo{author}{\bibfnamefont{M.}~\bibnamefont{Isi}} \bibnamefont{and} \bibinfo{author}{\bibfnamefont{W.~M.} \bibnamefont{Farr}} (\bibinfo{year}{2022}), \eprint{2202.02941}.
  
  \bibitem[{\citenamefont{Oshita}(2023)}]{Oshita:2022pkc}
  \bibinfo{author}{\bibfnamefont{N.}~\bibnamefont{Oshita}}, \bibinfo{journal}{JCAP} \textbf{\bibinfo{volume}{04}}, \bibinfo{pages}{013} (\bibinfo{year}{2023}), \eprint{2208.02923}.
  
  \bibitem[{\citenamefont{Fuller and Ma}(2019)}]{Fuller19}
  \bibinfo{author}{\bibfnamefont{J.}~\bibnamefont{Fuller}} \bibnamefont{and} \bibinfo{author}{\bibfnamefont{L.}~\bibnamefont{Ma}}, \bibinfo{journal}{Astrophys. J. Lett.} \textbf{\bibinfo{volume}{881}}, \bibinfo{pages}{L1} (\bibinfo{year}{2019}), \eprint{1907.03714}.
  
  \bibitem[{\citenamefont{Amaro-Seoane et~al.}(2017)}]{LISA17}
  \bibinfo{author}{\bibfnamefont{P.}~\bibnamefont{Amaro-Seoane}} \bibnamefont{et~al.} (\bibinfo{collaboration}{LISA}) (\bibinfo{year}{2017}), \eprint{1702.00786}.
  
  \bibitem[{\citenamefont{{Amaro-Seoane} et~al.}(2022)\citenamefont{{Amaro-Seoane}, {Andrews}, {Arca Sedda}, {Askar}, {Balasov}, {Bartos}, {Bavera}, {Bellovary}, {Berry}, {Berti} et~al.}}]{AmaroSeoane22}
  \bibinfo{author}{\bibfnamefont{P.}~\bibnamefont{{Amaro-Seoane}}}, \bibinfo{author}{\bibfnamefont{J.}~\bibnamefont{{Andrews}}}, \bibinfo{author}{\bibfnamefont{M.}~\bibnamefont{{Arca Sedda}}}, \bibinfo{author}{\bibfnamefont{A.}~\bibnamefont{{Askar}}}, \bibinfo{author}{\bibfnamefont{R.}~\bibnamefont{{Balasov}}}, \bibinfo{author}{\bibfnamefont{I.}~\bibnamefont{{Bartos}}}, \bibinfo{author}{\bibfnamefont{S.~S.} \bibnamefont{{Bavera}}}, \bibinfo{author}{\bibfnamefont{J.}~\bibnamefont{{Bellovary}}}, \bibinfo{author}{\bibfnamefont{C.~P.~L.} \bibnamefont{{Berry}}}, \bibinfo{author}{\bibfnamefont{E.}~\bibnamefont{{Berti}}}, \bibnamefont{et~al.}, \bibinfo{journal}{arXiv e-prints} \bibinfo{eid}{arXiv:2203.06016} (\bibinfo{year}{2022}), \eprint{2203.06016}.
  
  \bibitem[{\citenamefont{Dotti et~al.}(2013)\citenamefont{Dotti, Colpi, Pallini, Perego, and Volonteri}}]{Dotti13}
  \bibinfo{author}{\bibfnamefont{M.}~\bibnamefont{Dotti}}, \bibinfo{author}{\bibfnamefont{M.}~\bibnamefont{Colpi}}, \bibinfo{author}{\bibfnamefont{S.}~\bibnamefont{Pallini}}, \bibinfo{author}{\bibfnamefont{A.}~\bibnamefont{Perego}}, \bibnamefont{and} \bibinfo{author}{\bibfnamefont{M.}~\bibnamefont{Volonteri}}, \bibinfo{journal}{Astrophys. J.} \textbf{\bibinfo{volume}{762}}, \bibinfo{pages}{68} (\bibinfo{year}{2013}), \eprint{1211.4871}.
  
  \bibitem[{\citenamefont{Dubois et~al.}(2014)\citenamefont{Dubois, Volonteri, and Silk}}]{Dubois14}
  \bibinfo{author}{\bibfnamefont{Y.}~\bibnamefont{Dubois}}, \bibinfo{author}{\bibfnamefont{M.}~\bibnamefont{Volonteri}}, \bibnamefont{and} \bibinfo{author}{\bibfnamefont{J.}~\bibnamefont{Silk}}, \bibinfo{journal}{Mon. Not. Roy. Astron. Soc.} \textbf{\bibinfo{volume}{440}}, \bibinfo{pages}{1590} (\bibinfo{year}{2014}), \eprint{1304.4583}.
  
  \bibitem[{\citenamefont{Bustamante and Springel}(2019)}]{Bustamante19}
  \bibinfo{author}{\bibfnamefont{S.}~\bibnamefont{Bustamante}} \bibnamefont{and} \bibinfo{author}{\bibfnamefont{V.}~\bibnamefont{Springel}}, \bibinfo{journal}{Mon. Not. Roy. Astron. Soc.} \textbf{\bibinfo{volume}{490}}, \bibinfo{pages}{4133} (\bibinfo{year}{2019}), \eprint{1902.04651}.
  
  \bibitem[{\citenamefont{Barausse}(2012)}]{Barausse12}
  \bibinfo{author}{\bibfnamefont{E.}~\bibnamefont{Barausse}}, \bibinfo{journal}{Mon. Not. Roy. Astron. Soc.} \textbf{\bibinfo{volume}{423}}, \bibinfo{pages}{2533} (\bibinfo{year}{2012}), \eprint{1201.5888}.
  
  \bibitem[{\citenamefont{Reynolds}(2013)}]{Reynolds13}
  \bibinfo{author}{\bibfnamefont{C.~S.} \bibnamefont{Reynolds}}, \bibinfo{journal}{Class. Quant. Grav.} \textbf{\bibinfo{volume}{30}}, \bibinfo{pages}{244004} (\bibinfo{year}{2013}), \eprint{1307.3246}.
  
  \bibitem[{\citenamefont{Vasudevan et~al.}(2016)\citenamefont{Vasudevan, Fabian, Reynolds, Aird, Dauser, and Gallo}}]{Vasudevan15}
  \bibinfo{author}{\bibfnamefont{R.~V.} \bibnamefont{Vasudevan}}, \bibinfo{author}{\bibfnamefont{A.~C.} \bibnamefont{Fabian}}, \bibinfo{author}{\bibfnamefont{C.~S.} \bibnamefont{Reynolds}}, \bibinfo{author}{\bibfnamefont{J.}~\bibnamefont{Aird}}, \bibinfo{author}{\bibfnamefont{T.}~\bibnamefont{Dauser}}, \bibnamefont{and} \bibinfo{author}{\bibfnamefont{L.~C.} \bibnamefont{Gallo}}, \bibinfo{journal}{Mon. Not. Roy. Astron. Soc.} \textbf{\bibinfo{volume}{458}}, \bibinfo{pages}{2012} (\bibinfo{year}{2016}), \eprint{1506.01027}.
  
  \bibitem[{\citenamefont{Reynolds}(2021)}]{Reynolds21}
  \bibinfo{author}{\bibfnamefont{C.~S.} \bibnamefont{Reynolds}}, \bibinfo{journal}{Ann. Rev. Astron. Astrophys.} \textbf{\bibinfo{volume}{59}}, \bibinfo{pages}{117} (\bibinfo{year}{2021}), \eprint{2011.08948}.
  
  \bibitem[{\citenamefont{Berti et~al.}(2006)\citenamefont{Berti, Cardoso, and Will}}]{Berti:2005ys}
  \bibinfo{author}{\bibfnamefont{E.}~\bibnamefont{Berti}}, \bibinfo{author}{\bibfnamefont{V.}~\bibnamefont{Cardoso}}, \bibnamefont{and} \bibinfo{author}{\bibfnamefont{C.~M.} \bibnamefont{Will}}, \bibinfo{journal}{Phys. Rev. D} \textbf{\bibinfo{volume}{73}}, \bibinfo{pages}{064030} (\bibinfo{year}{2006}), \eprint{gr-qc/0512160}.
  
  \bibitem[{\citenamefont{Ota and Chirenti}(2022)}]{Ota:2021ypb}
  \bibinfo{author}{\bibfnamefont{I.}~\bibnamefont{Ota}} \bibnamefont{and} \bibinfo{author}{\bibfnamefont{C.}~\bibnamefont{Chirenti}}, \bibinfo{journal}{Phys. Rev. D} \textbf{\bibinfo{volume}{105}}, \bibinfo{pages}{044015} (\bibinfo{year}{2022}), \eprint{2108.01774}.
  
  \bibitem[{\citenamefont{Bhagwat et~al.}(2022)\citenamefont{Bhagwat, Pacilio, Barausse, and Pani}}]{Bhagwat:2021kwv}
  \bibinfo{author}{\bibfnamefont{S.}~\bibnamefont{Bhagwat}}, \bibinfo{author}{\bibfnamefont{C.}~\bibnamefont{Pacilio}}, \bibinfo{author}{\bibfnamefont{E.}~\bibnamefont{Barausse}}, \bibnamefont{and} \bibinfo{author}{\bibfnamefont{P.}~\bibnamefont{Pani}}, \bibinfo{journal}{Phys. Rev. D} \textbf{\bibinfo{volume}{105}}, \bibinfo{pages}{124063} (\bibinfo{year}{2022}), \eprint{2201.00023}.
  
  \bibitem[{\citenamefont{Sasaki and Nakamura}(1982)}]{Sasaki:1981sx}
  \bibinfo{author}{\bibfnamefont{M.}~\bibnamefont{Sasaki}} \bibnamefont{and} \bibinfo{author}{\bibfnamefont{T.}~\bibnamefont{Nakamura}}, \bibinfo{journal}{Prog. Theor. Phys.} \textbf{\bibinfo{volume}{67}}, \bibinfo{pages}{1788} (\bibinfo{year}{1982}).
  
  \bibitem[{\citenamefont{Kojima and Nakamura}(1984)}]{Kojima:1984cj}
  \bibinfo{author}{\bibfnamefont{Y.}~\bibnamefont{Kojima}} \bibnamefont{and} \bibinfo{author}{\bibfnamefont{T.}~\bibnamefont{Nakamura}}, \bibinfo{journal}{Prog. Theor. Phys.} \textbf{\bibinfo{volume}{71}}, \bibinfo{pages}{79} (\bibinfo{year}{1984}).
  
  \bibitem[{\citenamefont{Jaramillo et~al.}(2021)\citenamefont{Jaramillo, Panosso~Macedo, and Al~Sheikh}}]{Jaramillo:2020tuu}
  \bibinfo{author}{\bibfnamefont{J.~L.} \bibnamefont{Jaramillo}}, \bibinfo{author}{\bibfnamefont{R.}~\bibnamefont{Panosso~Macedo}}, \bibnamefont{and} \bibinfo{author}{\bibfnamefont{L.}~\bibnamefont{Al~Sheikh}}, \bibinfo{journal}{Phys. Rev. X} \textbf{\bibinfo{volume}{11}}, \bibinfo{pages}{031003} (\bibinfo{year}{2021}), \eprint{2004.06434}.
  
  \bibitem[{\citenamefont{Cardoso et~al.}(2016)\citenamefont{Cardoso, Franzin, and Pani}}]{Cardoso:2016rao}
  \bibinfo{author}{\bibfnamefont{V.}~\bibnamefont{Cardoso}}, \bibinfo{author}{\bibfnamefont{E.}~\bibnamefont{Franzin}}, \bibnamefont{and} \bibinfo{author}{\bibfnamefont{P.}~\bibnamefont{Pani}}, \bibinfo{journal}{Phys. Rev. Lett.} \textbf{\bibinfo{volume}{116}}, \bibinfo{pages}{171101} (\bibinfo{year}{2016}), \bibinfo{note}{[Erratum: Phys.Rev.Lett. 117, 089902 (2016)]}, \eprint{1602.07309}.
  
  \bibitem[{\citenamefont{Oshita and Afshordi}(2019)}]{Oshita:2018fqu}
  \bibinfo{author}{\bibfnamefont{N.}~\bibnamefont{Oshita}} \bibnamefont{and} \bibinfo{author}{\bibfnamefont{N.}~\bibnamefont{Afshordi}}, \bibinfo{journal}{Phys. Rev. D} \textbf{\bibinfo{volume}{99}}, \bibinfo{pages}{044002} (\bibinfo{year}{2019}), \eprint{1807.10287}.
  
  \bibitem[{\citenamefont{Finch and Moore}(2021)}]{Finch:2021qph}
  \bibinfo{author}{\bibfnamefont{E.}~\bibnamefont{Finch}} \bibnamefont{and} \bibinfo{author}{\bibfnamefont{C.~J.} \bibnamefont{Moore}}, \bibinfo{journal}{Phys. Rev. D} \textbf{\bibinfo{volume}{104}}, \bibinfo{pages}{123034} (\bibinfo{year}{2021}), \eprint{2108.09344}.
  
  \bibitem[{\citenamefont{Ferrari and Mashhoon}(1984)}]{Ferrari:1984zz}
  \bibinfo{author}{\bibfnamefont{V.}~\bibnamefont{Ferrari}} \bibnamefont{and} \bibinfo{author}{\bibfnamefont{B.}~\bibnamefont{Mashhoon}}, \bibinfo{journal}{Phys. Rev. D} \textbf{\bibinfo{volume}{30}}, \bibinfo{pages}{295} (\bibinfo{year}{1984}).
  
  \bibitem[{\citenamefont{Mourier et~al.}(2021)\citenamefont{Mourier, Jim\'enez~Forteza, Pook-Kolb, Krishnan, and Schnetter}}]{Mourier:2020mwa}
  \bibinfo{author}{\bibfnamefont{P.}~\bibnamefont{Mourier}}, \bibinfo{author}{\bibfnamefont{X.}~\bibnamefont{Jim\'enez~Forteza}}, \bibinfo{author}{\bibfnamefont{D.}~\bibnamefont{Pook-Kolb}}, \bibinfo{author}{\bibfnamefont{B.}~\bibnamefont{Krishnan}}, \bibnamefont{and} \bibinfo{author}{\bibfnamefont{E.}~\bibnamefont{Schnetter}}, \bibinfo{journal}{Phys. Rev. D} \textbf{\bibinfo{volume}{103}}, \bibinfo{pages}{044054} (\bibinfo{year}{2021}), \eprint{2010.15186}.
  
  \bibitem[{\citenamefont{Robson et~al.}(2019)\citenamefont{Robson, Cornish, and Liu}}]{Robson:2018ifk}
  \bibinfo{author}{\bibfnamefont{T.}~\bibnamefont{Robson}}, \bibinfo{author}{\bibfnamefont{N.~J.} \bibnamefont{Cornish}}, \bibnamefont{and} \bibinfo{author}{\bibfnamefont{C.}~\bibnamefont{Liu}}, \bibinfo{journal}{Class. Quant. Grav.} \textbf{\bibinfo{volume}{36}}, \bibinfo{pages}{105011} (\bibinfo{year}{2019}), \eprint{1803.01944}.
  
  \bibitem[{\citenamefont{Cabero et~al.}(2020)\citenamefont{Cabero, Westerweck, Capano, Kumar, Nielsen, and Krishnan}}]{Cabero20}
  \bibinfo{author}{\bibfnamefont{M.}~\bibnamefont{Cabero}}, \bibinfo{author}{\bibfnamefont{J.}~\bibnamefont{Westerweck}}, \bibinfo{author}{\bibfnamefont{C.~D.} \bibnamefont{Capano}}, \bibinfo{author}{\bibfnamefont{S.}~\bibnamefont{Kumar}}, \bibinfo{author}{\bibfnamefont{A.~B.} \bibnamefont{Nielsen}}, \bibnamefont{and} \bibinfo{author}{\bibfnamefont{B.}~\bibnamefont{Krishnan}}, \bibinfo{journal}{Phys. Rev. D} \textbf{\bibinfo{volume}{101}}, \bibinfo{pages}{064044} (\bibinfo{year}{2020}), \eprint{1911.01361}.
  
  \bibitem[{\citenamefont{Abbott et~al.}(2016{\natexlab{c}})}]{LIGOScientific:2016aoc}
  \bibinfo{author}{\bibfnamefont{B.~P.} \bibnamefont{Abbott}} \bibnamefont{et~al.} (\bibinfo{collaboration}{LIGO Scientific, Virgo}), \bibinfo{journal}{Phys. Rev. Lett.} \textbf{\bibinfo{volume}{116}}, \bibinfo{pages}{061102} (\bibinfo{year}{2016}{\natexlab{c}}), \eprint{1602.03837}.
  
  \bibitem[{\citenamefont{Miller}(2004)}]{Miller05}
  \bibinfo{author}{\bibfnamefont{M.~C.} \bibnamefont{Miller}}, \bibinfo{journal}{Astrophys. J.} \textbf{\bibinfo{volume}{618}}, \bibinfo{pages}{426} (\bibinfo{year}{2004}), \eprint{astro-ph/0409331}.
  
  \bibitem[{\citenamefont{Portegies~Zwart et~al.}(2006)\citenamefont{Portegies~Zwart, Baumgardt, McMillan, Makino, Hut, and Ebisuzaki}}]{PortegiesZwart06}
  \bibinfo{author}{\bibfnamefont{S.~F.} \bibnamefont{Portegies~Zwart}}, \bibinfo{author}{\bibfnamefont{H.}~\bibnamefont{Baumgardt}}, \bibinfo{author}{\bibfnamefont{S.~L.~W.} \bibnamefont{McMillan}}, \bibinfo{author}{\bibfnamefont{J.}~\bibnamefont{Makino}}, \bibinfo{author}{\bibfnamefont{P.}~\bibnamefont{Hut}}, \bibnamefont{and} \bibinfo{author}{\bibfnamefont{T.}~\bibnamefont{Ebisuzaki}}, \bibinfo{journal}{Astrophys. J.} \textbf{\bibinfo{volume}{641}}, \bibinfo{pages}{319} (\bibinfo{year}{2006}), \eprint{astro-ph/0511397}.
  
  \bibitem[{\citenamefont{Matsubayashi et~al.}(2007)\citenamefont{Matsubayashi, Makino, and Ebisuzaki}}]{Matsubayashi07}
  \bibinfo{author}{\bibfnamefont{T.}~\bibnamefont{Matsubayashi}}, \bibinfo{author}{\bibfnamefont{J.}~\bibnamefont{Makino}}, \bibnamefont{and} \bibinfo{author}{\bibfnamefont{T.}~\bibnamefont{Ebisuzaki}}, \bibinfo{journal}{Astrophys. J.} \textbf{\bibinfo{volume}{656}}, \bibinfo{pages}{879} (\bibinfo{year}{2007}), \eprint{astro-ph/0511782}.
  
  \bibitem[{\citenamefont{Arca-Sedda and Capuzzo-Dolcetta}(2019)}]{Arca-Sedda18}
  \bibinfo{author}{\bibfnamefont{M.}~\bibnamefont{Arca-Sedda}} \bibnamefont{and} \bibinfo{author}{\bibfnamefont{R.}~\bibnamefont{Capuzzo-Dolcetta}}, \bibinfo{journal}{Mon. Not. Roy. Astron. Soc.} \textbf{\bibinfo{volume}{483}}, \bibinfo{pages}{152} (\bibinfo{year}{2019}), \eprint{1709.05567}.
  
  \bibitem[{\citenamefont{Arca-Sedda and Gualandris}(2018)}]{Arca-Sedda19}
  \bibinfo{author}{\bibfnamefont{M.}~\bibnamefont{Arca-Sedda}} \bibnamefont{and} \bibinfo{author}{\bibfnamefont{A.}~\bibnamefont{Gualandris}}, \bibinfo{journal}{Mon. Not. Roy. Astron. Soc.} \textbf{\bibinfo{volume}{477}}, \bibinfo{pages}{4423} (\bibinfo{year}{2018}), \eprint{1804.06116}.
  
  \bibitem[{\citenamefont{Amaro-Seoane}(2018)}]{Amaro-Seoane18}
  \bibinfo{author}{\bibfnamefont{P.}~\bibnamefont{Amaro-Seoane}}, \bibinfo{journal}{Living Rev. Rel.} \textbf{\bibinfo{volume}{21}}, \bibinfo{pages}{4} (\bibinfo{year}{2018}), \eprint{1205.5240}.
  
  \bibitem[{\citenamefont{Bar-Or and Alexander}(2016)}]{Bar-Or15}
  \bibinfo{author}{\bibfnamefont{B.}~\bibnamefont{Bar-Or}} \bibnamefont{and} \bibinfo{author}{\bibfnamefont{T.}~\bibnamefont{Alexander}}, \bibinfo{journal}{Astrophys. J.} \textbf{\bibinfo{volume}{820}}, \bibinfo{pages}{129} (\bibinfo{year}{2016}), \eprint{1508.01390}.
  
  \bibitem[{\citenamefont{Babak et~al.}(2017)\citenamefont{Babak, Gair, Sesana, Barausse, Sopuerta, Berry, Berti, Amaro-Seoane, Petiteau, and Klein}}]{Babak18}
  \bibinfo{author}{\bibfnamefont{S.}~\bibnamefont{Babak}}, \bibinfo{author}{\bibfnamefont{J.}~\bibnamefont{Gair}}, \bibinfo{author}{\bibfnamefont{A.}~\bibnamefont{Sesana}}, \bibinfo{author}{\bibfnamefont{E.}~\bibnamefont{Barausse}}, \bibinfo{author}{\bibfnamefont{C.~F.} \bibnamefont{Sopuerta}}, \bibinfo{author}{\bibfnamefont{C.~P.~L.} \bibnamefont{Berry}}, \bibinfo{author}{\bibfnamefont{E.}~\bibnamefont{Berti}}, \bibinfo{author}{\bibfnamefont{P.}~\bibnamefont{Amaro-Seoane}}, \bibinfo{author}{\bibfnamefont{A.}~\bibnamefont{Petiteau}}, \bibnamefont{and} \bibinfo{author}{\bibfnamefont{A.}~\bibnamefont{Klein}}, \bibinfo{journal}{Phys. Rev. D} \textbf{\bibinfo{volume}{95}}, \bibinfo{pages}{103012} (\bibinfo{year}{2017}), \eprint{1703.09722}.
  
  \bibitem[{\citenamefont{Rose et~al.}(2022)\citenamefont{Rose, Naoz, Sari, and Linial}}]{Rose22}
  \bibinfo{author}{\bibfnamefont{S.~C.} \bibnamefont{Rose}}, \bibinfo{author}{\bibfnamefont{S.}~\bibnamefont{Naoz}}, \bibinfo{author}{\bibfnamefont{R.}~\bibnamefont{Sari}}, \bibnamefont{and} \bibinfo{author}{\bibfnamefont{I.}~\bibnamefont{Linial}}, \bibinfo{journal}{Astrophys. J. Lett.} \textbf{\bibinfo{volume}{929}}, \bibinfo{pages}{L22} (\bibinfo{year}{2022}), \eprint{2201.00022}.
  
  \end{thebibliography}
\end{document}